\documentclass[aps,prx,twocolumn,
			   groupedaddress,superscriptaddress,
			   amsfonts,amssymb,amsmath,
			   citeautoscript,longbibliography,
			   a4paper
			   ]{revtex4-2}

\usepackage[utf8]{inputenc}
\usepackage[english]{babel}

\usepackage{microtype} 
\usepackage{xspace} 

\usepackage{txfonts}  
\usepackage{txfontsb} 

\usepackage{bm} 

\usepackage{xcolor}
\usepackage[]{graphicx} 
\graphicspath{{figs/}}

\usepackage[]{booktabs}
\usepackage{array}
\usepackage{layouts}
\usepackage{multirow}

\usepackage{enumerate}
\usepackage[inline]{enumitem}

\usepackage{hyperref}
\hypersetup{colorlinks,
    linkcolor=black,
    citecolor=black,
    urlcolor=black
}

\usepackage[capitalize,nameinlink]{cleveref}

\crefname{subequations}{Eqs.}{Eqs.} 
\Crefname{subequations}{Eqs.}{Eqs.}
\crefformat{subequations}{#2Eqs.~(#1)#3}
\Crefformat{subequations}{#2Eqs.~(#1)#3}
\crefname{page}{p.}{p.} 

\usepackage{placeins}

\usepackage{siunitx}
\DeclareSIUnit[number-unit-product = ]\percent{\char`\%} 

\usepackage[centering,hmargin=16mm,tmargin=30mm,bmargin=26mm]{geometry}

\usepackage{soul}

\usepackage{textcomp} 
\usepackage{xifthen}
\usepackage{etoolbox}
\newboolean{togglecomments}
\newboolean{togglechanges}

\setboolean{togglecomments}{true}
\setboolean{togglechanges}{false}

\newcommand{\textblacksquare}{$\blacksquare$}
\newcommand{\todo}[1]{\ifbool{togglecomments}%
	{\textcolor{green!60!black}{\small\textsf{{}\textsuperscript{\textsc{\textsf{todo}}}}[#1]}} 
	{}}     
\newcommand{\comment}[2]{\ifbool{togglecomments}%
		{\textcolor{blue!70!black}{\small\sf\textsuperscript{\textsc{\textsf{#1}}}[#2]}} 
		{}}     
\newcommand{\swap}[2]{\ifbool{togglechanges}
	{#2}  
	{\textcolor{red!70!black}{[\ignorespaces#1]}\textrightarrow{}\textcolor{green!50!black}{[\ignorespaces#2]}}}
\newcommand{\remove}[1]{\ifbool{togglechanges}
	{}    
	{\textcolor{red!70!black}{\ignorespaces#1}}}
\newcommand{\inset}[1]{\ifbool{togglechanges}
	{#1}  
	{\textcolor{green!50!black}{#1}}}
\newcommand{\citeremind}[1]{%
	[\textcolor{blue!75!black!80!yellow}{\textblacksquare%
		\ifthenelse{\isempty{#1}}{}{\textsuperscript{\tiny\textsf{#1}}}%
	}]\xspace}

\newcommand{\appropto}{\mathrel{\vcenter{
			\offinterlineskip\halign{\hfil$##$\cr
				\propto\cr\noalign{\kern.2pt}\sim\cr\noalign{\kern-2.5pt}}}}}

\makeatletter
\newcommand{\raisemath}[1]{\mathpalette{\raisem@th{#1}}}
\newcommand{\raisem@th}[3]{\raisebox{#1}{$#2#3$}}
\makeatother

\renewcommand{\paragraph}[1]{\vskip 1ex\noindent\textbf{#1.}~}

\usepackage[eulergreek]{sansmath}
\makeatletter
\renewcommand\@make@capt@title[2]{%
    \@ifx@empty\float@link{\@firstofone}{\expandafter\href\expandafter{\float@link}}%
    \sisetup{math-sf=\textsf}%
    \sansmath\sffamily\textbf{#1\@caption@fignum@sep}#2 
}%

\makeatother

\newcommand{\mitphysicsaffil}{\footnotesize Department of Physics, Massachusetts Institute of Technology, Cambridge, Massachusetts 02139, USA}
\newcommand{\miteecsaffil}{\footnotesize Department of Electrical Engineering and Computer Science, Massachusetts Institute of Technology, Cambridge, Massachusetts 02139, USA}

\begin{document}

\author{Andrew~Ma}
\affiliation{\miteecsaffil}
\author{Marin~Solja\v{c}i\'{c}}
\email{soljacic@mit.edu}
\affiliation{\mitphysicsaffil}

\title{Learning simple heuristic rules for classifying materials based on chemical composition}

\begin{abstract}
    \noindent In the past decade, there has been a significant interest in the use of machine learning approaches in materials science research.  Conventional deep learning approaches that rely on complex, nonlinear models have become increasingly important in computational materials science due to their high predictive accuracy.  In contrast to these approaches, we have shown in a recent work that a remarkably simple learned heuristic rule -- based on the concept of topogivity -- can classify whether a material is topological using only its chemical composition.  In this paper, we go beyond the topology classification scenario by also studying the use of machine learning to develop simple heuristic rules for classifying whether a material is a metal based on chemical composition.  Moreover, we present a framework for incorporating chemistry-informed inductive bias based on the structure of the periodic table.  For both the topology classification and the metallicity classification tasks, we empirically characterize the performance of simple heuristic rules fit with and without chemistry-informed inductive bias across a wide range of training set sizes.  We find evidence that incorporating chemistry-informed inductive bias can reduce the amount of training data required to reach a given level of test accuracy.
\end{abstract}

\maketitle

\section{Introduction}

Data-driven approaches have been attracting increasing interest in the field of materials science, demonstrating potential both in improving our scientific understanding of materials as well as in driving the discovery of new materials~\cite{sun2019map,deringer2021origins,zhong2020accelerated,oliynyk2016high,yao2021inverse,himanen2019data,saal2020machine}.
Enabled in part by the expanding amount of available materials data~\cite{hellenbrandt2004inorganic,curtarolo2012aflow,jain2013commentary,saal2013materials,kim2020band,dunn2020benchmarking,tang2019comprehensive,zhang2019catalogue,vergniory2019complete,vergniory2022all}, a wide variety of machine learning (ML) methods have been developed for materials, including models that use both chemical composition and structural information~\cite{xie2018crystal,schutt2018schnet,chen2021direct,batzner20223,xiao2023invertible,ward2017including,choudhary2018machine} as well as models that use only chemical composition information~\cite{jha2018elemnet,goodall2020predicting,ward2016general,legrain2017chemical,zhang2021finding}.
While ab initio methods are still typically more accurate for modeling materials, ML methods are becoming a valuable part of the computational toolkit due to their computational speed as well as their ability to reveal insights from trends in data~\cite{schleder2019dft,axelrod2022learning,huang2023central}.

In the study of molecules and materials, chemical heuristic approaches have long played an important role~\cite{seeman2022understanding,karen2023heuristic,pauling1929principles,allen1993van}.
While the development of quantum mechanical approaches has enabled a more complete and faithful picture of molecules and materials, chemical heuristics still remain valuable because they give quick answers and provide valuable intuition.
In the modern materials informatics era, there has been a recent interest in both evaluating existing chemical heuristics and developing new chemical heuristics using data-driven approaches for materials~\cite{george2020chemist,george2020limited,filip2018geometric,bartel2019new,liu2023materials}.
From the point of view of ML for materials science, heuristic chemical models developed via ML represent a part of the broader landscape of interpretable ML for materials, which is an important area of research featuring a number of different approaches~\cite{oviedo2022interpretable,allen2022machine,muckley2023interpretable,chen2022accurate,isayev2017universal,ouyang2018sisso,zhang2023physically,ziletti2018insightful,moro2025multimodal,xu2024predicting,cao2020artificial,liu2021screening,ouyang2019simultaneous}.

In contemporary condensed matter, electronic band topology is a highly active area of research~\cite{bansil2016colloquium}.
Recently, in \citet{ma2023topogivity}, we introduced the topogivity approach, which constitutes a heuristic chemical rule approach for diagnosing whether a given material is topological.
In contrast to technically challenging ab initio diagnosis methods~\cite{xiao2021first}, our heuristic chemical rule was remarkably simple and provided valuable chemical intuition.
Our approach also contrasts with the typically much more complicated models that have been used in other broadly applicable ML approaches for topology diagnosis in real materials~\cite{claussen2020detection,andrejevic2022machine,rasul2024machine,xu2024predicting}.
Moreover, we demonstrated the power of this rule by applying it to enable the discovery of new topological materials that are not detectable by standard symmetry-based approaches~\cite{fu2007topologicalinv,po2017symmetry,bradlyn2017topological,slager2017,song2018quantitative}.
Subsequently, other groups have built on this work.
Ref.~\cite{he2025machine} fitted a topogivity-based model on a more comprehensive dataset, as part of an extensive study on ML for topological materials that included both interpretable and black-box models.
Ref. \cite{hong2025discovery} integrated the topogivity model into a pipeline for topological materials discovery that also involves the use of generative models.
Expanding on this direction of work, in this paper, we take up a broader study of the style of modeling that we introduced in our topogivity work~\cite{ma2023topogivity}.

In ML terms, the topology diagnosis task considered in our topogivity work~\cite{ma2023topogivity} was a binary classification question (note that topology diagnosis is sometimes treated as a task with more than two classes in other ML works).
Another binary classification question in ML for materials is the task of predicting whether or not a material is a metal, which has been studied using many different approaches of varying levels of accuracy and interpretability~\cite{xie2018crystal,ouyang2018sisso,isayev2017universal,jha2019enhancing,ouyang2019simultaneous,choudhary2018machine,zhuo2018predicting,wang2022learning,jung2024automatic,dunn2020benchmarking,zhou2018learning}.
Moreover, just as there is chemical intuition regarding electronic band topology~\cite{kumar2020topological,schoop2018chemical,khoury2021chemical,gibson2015three,isaeva2020crystal,klemenz2020role,claussen2020detection,liu2023materials}, it is even more well known -- at the textbook chemistry level in fact -- that there is chemical intuition regarding whether a material is a metal.
This makes the metallicity classification task a potentially suitable setting for applying the style of modeling developed in our topogivity work.
One of the contributions of this paper is the development of heuristic chemical rules for metallicity classification.

An important question in developing ML models for materials is the choice of material representation -- this choice determines what the model can directly use as input and may often explicitly or implicitly capture information about the degree of similarity between different elements~\cite{damewood2023representations,onwuli2023element}.
The choice of representation can often be important for the incorporation of physics and chemistry knowledge as inductive biases into ML models -- the incorporation of such knowledge is a major theme in ML for physical sciences more broadly~\cite{battaglia2018relational,karniadakis2021physics,childs2019embedding}.
A methodological development that we present in this work is a framework for incorporating chemistry-informed inductive bias, which can be understood in terms of designing a material representation that builds in the structure of the periodic table.
Equivalently, as we will see, this framework can also be understood in terms of weight tying.

In Section~\ref{section_modeling_framework}, we mathematically describe the modeling framework.
The first part of this section covers what we will term a full model, which is the model formulation based on one parameter per element that was originally introduced in our topogivity work~\cite{ma2023topogivity}.
The second part of this section covers the framework for incorporating chemistry-informed inductive bias to create restricted models and describes the relationship of restricted models to full models.
In Section~\ref{section_empirical_results}, we present empirical results.
This includes an extensive evaluation of both the full model and and the restricted model approaches on both the topology classification and metallicity classification tasks across a wide range of training set sizes.
This extensive evaluation provides evidence that the amount of training data required to achieve a given level of test accuracy can be lower for the restricted model than for the full model.
The empirical results also include an interpretable visualization of the full model for the metallicity classification task.
We close in Section~\ref{section_discussion} with a discussion of our findings and their significance.

\section{Modeling Framework \label{section_modeling_framework}}

In this section, we present our modeling framework, which includes two types of models.
The problem setting is a supervised learning, binary classification setting where we have materials $M$ as well as labels $y \in \{-1,1\}$.
The goal is to develop a classifier $\hat{y}(M)$ that can predict the label $y$ given an input material $M$.
The information that the models explicitly use as input for a given material is contained entirely in the chemical composition of the material, as specified by the collection of element fractions $\{ f_E(M)\}_{E \in \Omega}$, where $f_E(M)$ is the fraction of atoms in the unit cell of material $M$ that are element $E$ and $\Omega$ is the set of all elements present in the dataset.
For example, for the material $A_x B_y C_z$, we have $f_A(A_x B_y C_z) = \frac{x}{x+y+z}$, $f_B(A_x B_y C_z) = \frac{y}{x+y+z}$, $f_C(A_x B_y C_z) = \frac{z}{x+y+z}$, and $f_E(A_x B_y C_z) = 0 \ \forall E \ \notin \{A, B, C\}$.

\begin{figure*}[!htb]
    \centering
    \includegraphics[scale=0.67]{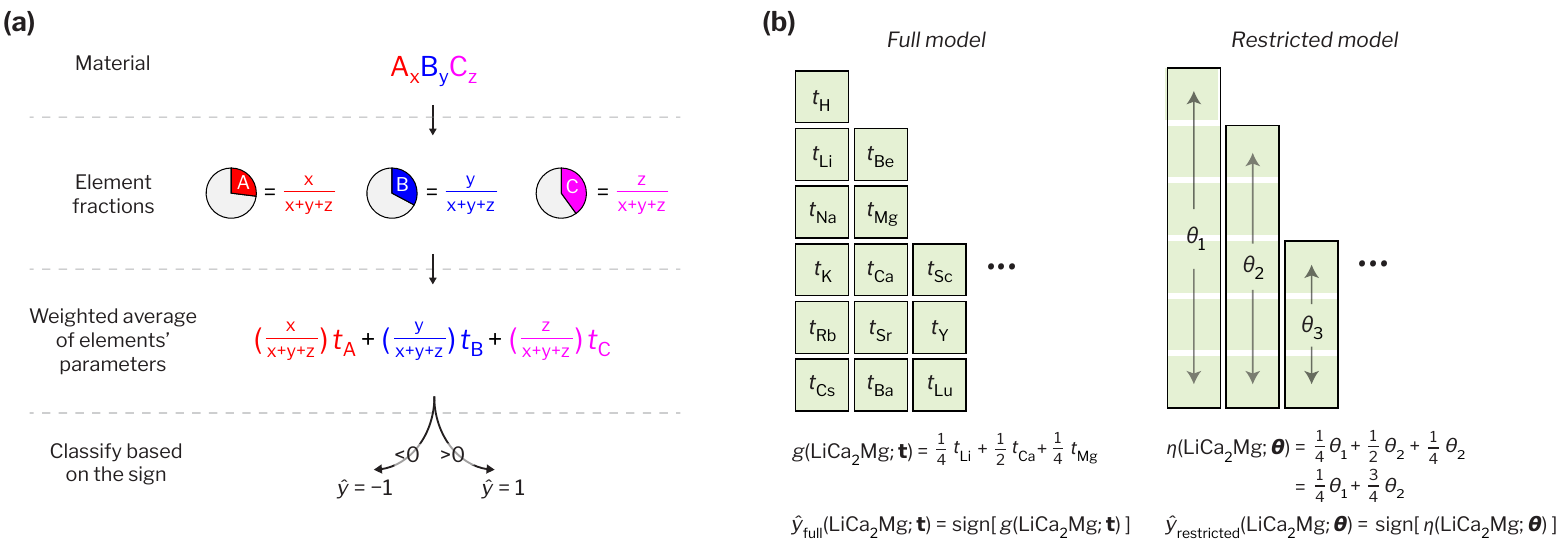}
    \caption{\textbf{Overview of modeling framework}. \textbf{(a)} For a given material, the simple heuristic rules predict a binary label using just the element fractions by taking the sign of a weighted average of the parameters of the constituent elements.
    \textbf{(b)} Both the full model and the restricted model can be understood from this weighted average of parameters perspective -- the full model has a separate parameter for each element whereas the restricted model involves parameter tying so that multiple elements share the same parameter.  The parameter tying in the restricted model approach can enable the incorporation of chemistry-informed inductive bias.  Each model's approach for diagnosis is illustrated with the example material $\mathrm{Li Ca_2 Mg}$.
    \label{fig_overview_schematic_of_modeling_framework}}
\end{figure*}

For both types of model, the model will take the form of a simple heuristic rule -- given the chemical composition, one needs only to look up the values of the elements' parameters in a table and then look at the sign of an appropriately weighted average to obtain the diagnosis.
The first type of model, which we term the full model, has exactly one parameter for each element; it is described in Section~\ref{subsection_full_model}.
The second type of model, which we term the restricted model, incorporates chemistry-informed inductive bias and can be understood as restricting the hypothesis class of the full model; it is described in Section~\ref{subsection_restricted_model}.
An overview schematic for both models is shown in Fig.~\ref{fig_overview_schematic_of_modeling_framework}.

The modeling framework described here is compatible with many different methods for learning parameters from data.  The actual training approach we use in our empirical work will be discussed in Section~\ref{section_empirical_results}; the current section is focused on the form of the models.

\subsection{Full Model \label{subsection_full_model}}

In this subsection, we describe the full model, which was originally introduced in our topogivity work in the specific context of diagnosing electronic topology~\cite{ma2023topogivity}.
Note that the notation in this paper will be slightly different from that of ref.~\cite{ma2023topogivity}, and so this subsection also serves to establish notation for the remainder of this paper.
Also note that while the mathematical form of the model in this paper is the same as in the one in ref.~\cite{ma2023topogivity}, the method we use for learning parameters from data in the empirical section of this work will not be identical to the methods in ref.~\cite{ma2023topogivity}.

Let $\mathbf{f}(M)$ be a vector containing all of the element fractions.
Let $t_E$ be a parameter for each element $E$ and let $\mathbf{t}$ be a vector containing all of these parameters (ordered so that the $i$-th entry of $\mathbf{t}$ and the $i$-th entry of $\mathbf{f}(M)$ correspond to the same element; e.g., in our code, we use the convention that both are ordered according to ascending atomic number).

We define the parameterized function $g(\cdot;\mathbf{t})$ as
\begin{equation}
    g(M;\mathbf{t}) = \sum_{E\in\Omega} t_E f_E(M) = \mathbf{t} \cdot \mathbf{f}(M),
\end{equation}
which can be viewed as a weighted average of the element parameters $t_E$ for all elements that are present in the material, where the weighting is with respect to the chemical formula subscripts.
Classification decisions are given by the sign of $g(M; \mathbf{t})$, i.e.,
\begin{equation}
    \hat{y}_{\mathrm{full}}(M; \mathbf{t}) = \mathrm{sign}[g(M; \mathbf{t})]. \label{eq_def_y_hat_full}
\end{equation}

For each element $E$, the parameter $t_E$ is intended to loosely capture the tendency of the element to form materials of the $y=+1$ class.
A greater value corresponds roughly to a greater tendency; note, however, that the classification rule is unchanged if all the parameters are scaled by a positive constant.
In this sense, the model provides heuristic chemical interpretability.
In ref.~\cite{ma2023topogivity}, the $y=+1$ class is the topological class, and so we termed the parameter $t_E$ the topogivity of element $E$; an in-depth discussion of the chemical picture provided by the notion of topogivity -- including its potential limitations -- is provided in ref.~\cite{ma2023topogivity}.

\subsection{Restricted Model \label{subsection_restricted_model}}

In this subsection, we introduce a framework for incorporating chemistry-informed inductive bias based on grouping together similar elements.
Our approach incorporates ideas from two different areas in the existing literature.
First, in the ML for materials science literature, there have been many works on models that build in knowledge regarding which elements are similar to each other, such as approaches based on hand-crafted representations incorporating domain knowledge from periodic table location or approaches utilizing learned vectors that capture element similarity~\cite{zheng2018machine,feng2021general,zhang2023physically,zhou2018learning,onwuli2023element}.
Periodic table information (such as row and column) has also been used in featurizing elements in graph neural networks (e.g., see ref. \cite{xie2018crystal}).
Second, outside of the domain of materials science, in the broader literature on linear models, there has been substantial work on developing methods that make parameters for different features take on the same value, either by using regularization to encourage parameter equality (e.g., penalizing the difference between parameters) or by tying parameter values together through aggregating features (by defining new super features that are each the sum or average of a group of features) so that it is guaranteed that features in the same group have the same value~\cite{tibshirani2005sparsity,bondell2008simultaneous,she2010sparse,park2007averaged,surer2021coefficient,surer2021glm_coefficient}.
In our approach, we integrate these two ideas by utilizing linear model feature tying to incorporate chemistry-based domain knowledge as an inductive bias.

We now describe in detail our framework for incorporating chemistry-informed inductive bias.
Let $P$ be a partition of $\Omega$.  $P$ then is a set where each member of the set is a set of elements, and we denote it as
\begin{equation}
    P = \{\omega_1, \omega_2, ..., \omega_{|P|} \}.
\end{equation}
For $i \in \{1, 2,..., |P|\}$, we define
\begin{equation}
    \nu_i(M) = \sum_{E\in\omega_i} f_E(M).\label{eq_def_nu_i_of_M}
\end{equation}
We denote the full vector of $\nu_i(M)$ as
\begin{equation}
    \boldsymbol{\nu}(M) = [\nu_1(M), \nu_2(M), ..., \nu_{|P|}(M)]^T .\label{eq_vector_nu_definition}
\end{equation}

Let $\boldsymbol{\theta}$ be the $|P|$-dimensional vector of parameters,
\begin{equation}
    \boldsymbol{\theta} = [\theta_1,\theta_2, ...,\theta_{|P|}]^T .
\end{equation}
We define $\eta(\cdot; \boldsymbol{\theta})$ as a function parameterized by $\boldsymbol{\theta}$, 
\begin{equation}
    \eta(M; \boldsymbol{\theta}) = \boldsymbol{\theta} \cdot \boldsymbol{\nu}(M) = \sum_{i=1}^{|P|} \theta_i \nu_i(M). \label{eq_def_eta_of_M}
\end{equation}
As in the full model, the classification decision is based on the sign, i.e.,
\begin{equation}
    \hat{y}_{\mathrm{restricted}}(M; \boldsymbol{\theta}) = \mathrm{sign}[\eta(M; \boldsymbol{\theta})]. \label{eq_def_y_hat_restricted}
\end{equation}

To understand the physical meaning of the $\eta(\cdot;\boldsymbol{\theta})$ formulation, we can observe the following relationship:
\begin{equation}
    \eta(\cdot; \boldsymbol{\theta}) = g(\cdot; [\theta_{\alpha(E_1)}, \theta_{\alpha(E_2)}, ..., \theta_{\alpha(E_{|\Omega|})}]^T), \label{eq_equiv_feature_aggregation_and_parameter_tying}
\end{equation}
where
\begin{equation}
    \alpha(E) = \mathrm{the} \ i \mathrm{\ that \ satisfies \ E \in \omega_i},
\end{equation}
and where we have numbered the elements such that $E_i$ is the element corresponding to the $i$-th entry of $\mathbf{f}(M)$, i.e.,
\begin{equation}
    \mathbf{f}(M) = [f_{E_1}(M),f_{E_2}(M), ...,f_{E_{|\Omega|}}(M)]^T.
\end{equation}
Note that Eq.~\eqref{eq_equiv_feature_aggregation_and_parameter_tying} immediately implies
\begin{equation}
    \hat{y}_{\mathrm{restricted}}(\cdot; \boldsymbol{\theta}) = \hat{y}_{\mathrm{full}}(\cdot;[\theta_{\alpha(E_1)}, \theta_{\alpha(E_2)}, ..., \theta_{\alpha(E_{|\Omega|})}]^T),
\end{equation}
which can be seen from applying the sign function to both sides of Eq.~\eqref{eq_equiv_feature_aggregation_and_parameter_tying} and then using Eqs. \eqref{eq_def_y_hat_full} and \eqref{eq_def_y_hat_restricted}.

The left hand side of Eq.~\ref{eq_equiv_feature_aggregation_and_parameter_tying} represents feature aggregation, which can be seen from the fact that $\eta(\cdot;\boldsymbol{\theta})$ acts on the vector of aggregated features.
The right hand side of Eq.~\ref{eq_equiv_feature_aggregation_and_parameter_tying} represents parameter tying; this can be seen from observing that
\begin{equation}
    \forall i \ \mathrm{such \ that} \ E_i \in \omega_j, \ \theta_{\alpha(E_i)} = \theta_j,
\end{equation}
which means that all of the elements in the same $\omega_j$ will have the same parameter value multiplying its associated element fraction in the $g(\cdot;\mathbf{t})$ formulation, and that parameter value will be $\theta_j$.
As such, the relationship in Eq.~\ref{eq_equiv_feature_aggregation_and_parameter_tying} represents an equivalence between parameter tying and summing-based feature aggregation, which is an equivalence that is sometimes utilized in the broader literature on linear models (e.g., see refs. \cite{park2007averaged,surer2021coefficient,surer2021glm_coefficient}); in our notation, the reason that this equivalence holds can easily be seen by substituting Eq.~\ref{eq_def_nu_i_of_M} into Eq.~\ref{eq_def_eta_of_M} and then manipulating terms, which yields $\eta(M; \boldsymbol{\theta}) = \sum_{i=1}^{|P|} \theta_i \sum_{E\in\omega_i} f_E(M) = \sum_{i=1}^{|P|}  \sum_{E\in\omega_i} \theta_{\alpha(E)} f_E(M) = \sum_{E \in \Omega} \theta_{\alpha(E)} f_E(M) = g(M; [\theta_{\alpha(E_1)}, \theta_{\alpha(E_2)}, ..., \theta_{\alpha(E_{|\Omega|})}]^T)$.

Every learnable function $\eta(\cdot;\boldsymbol{\theta})$ corresponds to some function $g(\cdot;\mathbf{t})$, but not vice versa.
From this, we can see the perspective that the relationship between the restricted model approach ($\eta(\cdot;\boldsymbol{\theta})$) and the full model approach ($g(\cdot;\mathbf{t})$) is that using the former can be understood as a way of restricting the hypothesis class to those where elements in the same member of the partition have equal values for their associated parameter.

This framework can in principle be used for any choice of partition $P$, but our main intention in developing this framework is for it to be used with choices of $P$ that appropriately build in domain knowledge.
Specifically, one can choose $P$ such that elements that one believes should behave similarly (for the task in mind) are grouped in the same subset $\omega_i$.
For all of the empirical results that we will show in this paper, we will take the choice that the subsets of $P$ correspond to the columns of the periodic table, plus one additional subset which is all of the f-block elements (see Supplementary Information for details).
The reason for this choice is that the periodic table is structured in a way such that elements in the same column (block) tend to be more similar to each other than to elements in other columns (blocks).
Tying together the weights of elements that we expect to behave similarly on the basis of the human understanding of chemistry reflected in the periodic table thus corresponds to a chemistry-informed inductive bias.
This is visualized for the first three columns of the periodic table in Fig.~\ref{fig_overview_schematic_of_modeling_framework}(b).
Note that this visualization does not show Francium (Fr), Radium (Ra), and Lawrencium (Lr) (which are elements in the first, second, and third columns, respectively), because those elements do not appear in any materials in either of the datasets used in the empirical portion of this paper (but in principle there is nothing preventing the modeling framework from being applied to those elements as well).

\section{Empirical Results \label{section_empirical_results}}

\subsection{Description of the Datasets \label{subsection_description_of_the_datasets}}

We consider two binary classification tasks.  The first is topology classification, for which the two classes are ``topological'' ($y=+1$) and ``trivial'' ($y=-1$).  The second is metallicity classification, for which the two classes are ``metal'' ($y=+1$) and ``not metal'' ($y=-1$).
We are in a supervised learning setting, so we need labeled datasets consisting of pairs of materials and their binary labels.

For the topology classification task, we use a labeled dataset with 9,026 materials, of which $49.0 \%$ are labeled as ``topological''.
This labeled dataset is the same as the labeled dataset that was used in ref.~\cite{ma2023topogivity}, which is a processed dataset that was constructed from the ab initio data generated in \citet{tang2019comprehensive}.
See ref.~\cite{ma2023topogivity} for further details on this topology dataset.
For the metallicity classification task, we use a labeled dataset with 4,898 materials, of which $49.6 \%$ are labeled as ``metal''.
To obtain this dataset, we slightly process one of the datasets from Matbench~\cite{dunn2020benchmarking}.
This Matbench dataset was originally curated from \citet{zhuo2018predicting}, which and it contains both experimental and ab initio data.
Details on our data processing procedure for the metallicity dataset are provided in the Supplementary Information.

As is the case with many materials datasets, both the topology dataset and the metallicity dataset that we use should be considered to have noisy labels, and this will impact the modeling results.
For the dataset that we use for the topology classification task, we provided a discussion of the sources of noise in ref.~\cite{ma2023topogivity}; additionally, a recent study helps to characterize potential systematic error in standard ab initio databases for electronic topology~\cite{mirhosseini2025revisiting}.

\subsection{Model Implementation and Training \label{subsection_model_implementation_and_training}}

We implement the full model by featurizing each material as a vector of element fractions $\mathbf{f}(M)$ and learning a linear binary classifier with no intercept to get the parameter vector $\mathbf{t}$.
As discussed in Section~\ref{subsection_restricted_model}, we can equivalently view the restricted model either in terms of using a representation based on feature aggregation or in terms of parameter tying.  For implementation purposes, we make use of the former viewpoint -- each material is featurized as a vector $\boldsymbol{\nu}(M)$ as defined in Eq.~\eqref{eq_vector_nu_definition}, and the parameter vector $\boldsymbol{\theta}$ is obtained by learning a linear binary classifier with no intercept.  As described in Section~\ref{subsection_restricted_model}, the choice of partition for the empirical results is based on periodic table columns and the f-block.

For all of the empirical results presented in this paper -- including both those for full model and those for the restricted model -- training is done by using weakly regularized logistic regression, which we implement using scikit-learn~\cite{pedregosa2011scikit}.
We use L2 regularization and set the inverse of regularization strength (as defined in the sci-kit learn documentation) to $C = 10^{6}$.  Further details are provided in the Supplementary Information.

\subsection{Performance Evaluation \label{subsection_performance_evaluation}}

In this subsection, we evaluate model performance on two tasks -- topology classification and metallicity classification -- by making use of the datasets described in Section~\ref{subsection_description_of_the_datasets}.
For each task, we evaluate the train and test performance of both the full model and the restricted model across a wide range of training set sizes.

For each task, the performance evaluation is performed using random training and test sets.
We consider 25 different training set sizes between 20 samples to 4,000 samples inclusive (on a log scale, the sizes are evenly spaced).
Test set size is fixed at 400 samples.
At each training set size, we generate a large number of pairs of training set and test set.
For each pair of training set and test set, the members of the sets are randomly selected without replacement from the overall dataset in a way such that the test set does not contain any materials with one or more elements that are not present among the training set materials (this method of selecting test sets is likely more favorable to the full model than some other reasonable methods for test set selection).
Due to the fact that there is typically greater variability in the test performance at smaller training set sizes, we use a greater number of random pairs of training set and test set for the smaller training set sizes.
Specifically, for a training set size $N_{\mathrm{train}}$, we generate $\frac{4 \times 10^5}{N_{\mathrm{train}}}$ pairs of training set and test set (e.g., this means that there are 20,000 pairs of training set and test set for the 20 training data points size and 100 pairs of training set and test set for the 4,000 training data points size).
For each pair of randomly selected training set and test set, we train and test both a full model as well as a restricted model.
For a given type of model and given training set size, results are aggregated across the many splits by taking the mean and standard deviation of each metric (train accuracy and test accuracy).
See Supplementary Information for further details.

\begin{figure}[!htb]
    \centering
    \includegraphics[width=0.5\textwidth]{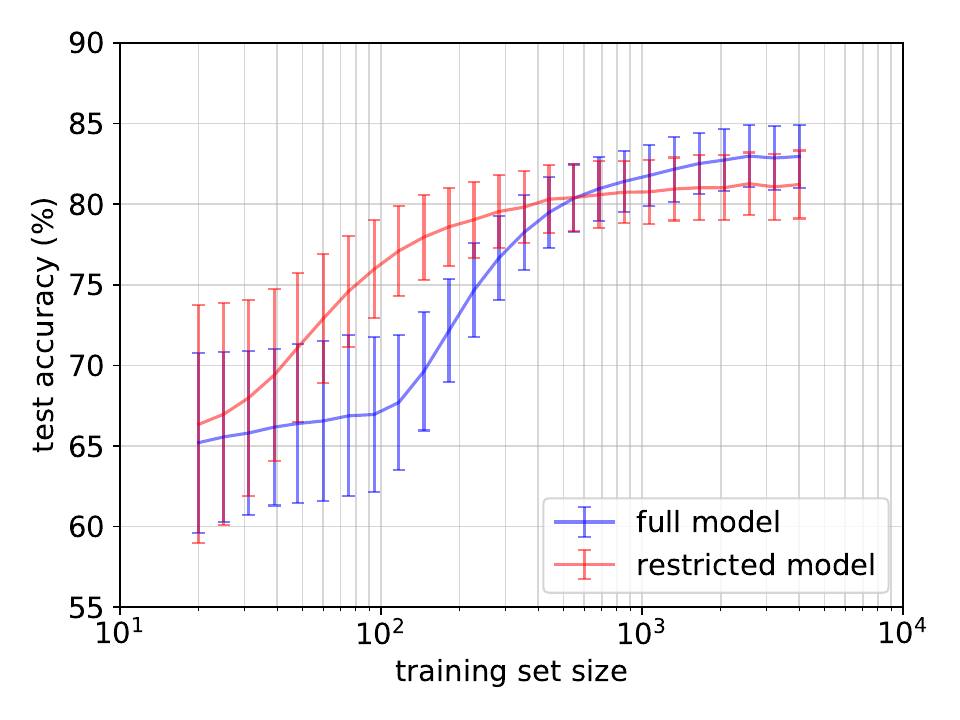}
    \includegraphics[width=0.5\textwidth]{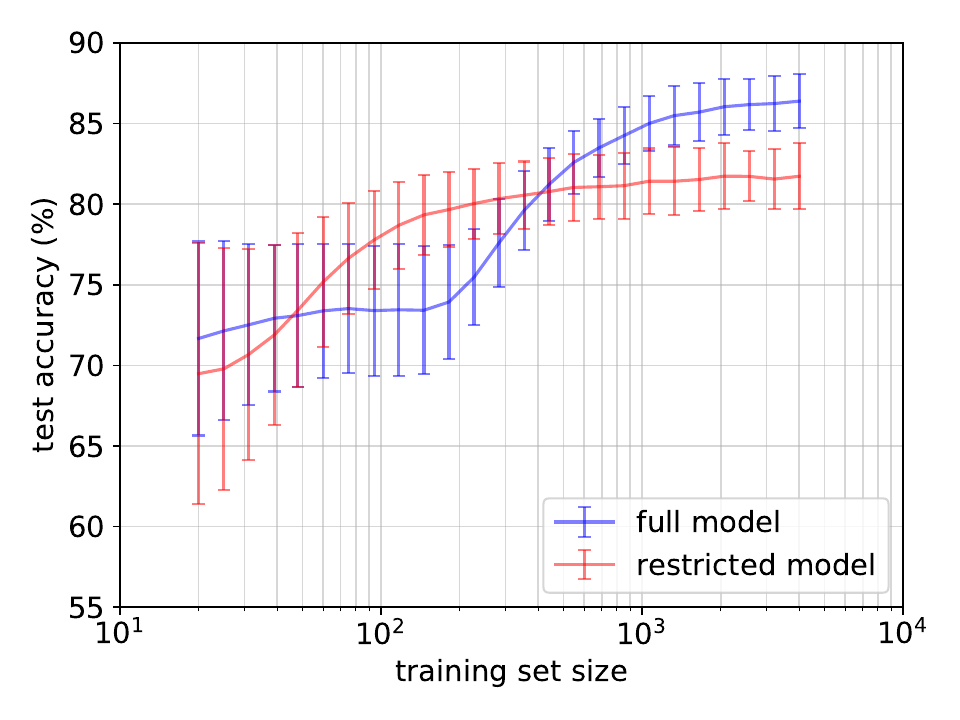}
    \caption{\textbf{Performance versus training set size.} The top plot and bottom plot respectively correspond to the the topology classification and metallicity classification tasks. For each plot, the mean and standard deviation of the test accuracy is shown for the full model and the restricted model at 25 different training set sizes.
    These curves were obtained by aggregating results across many iterations of train and test at each training set size (with more iterations for smaller training set sizes).  Test set size was fixed at 400 samples.
    \label{fig_test_learning_curves}}
\end{figure}

The models are implemented and trained using the procedure described in Section~\ref{subsection_model_implementation_and_training}.
We note, however, that if the goal were to simply get the best possible model at a given training set size, it may be possible to improve the performance of the full model and/or the restricted model by sweeping the regularization hyperparameter and/or trying other learning algorithms besides logistic regression.
Additionally, it may be possible to improve the restricted model performance by trying different partitions (in our empirical study in this paper we always take the choice where we partition based on periodic table columns and f-block).
These types of detailed explorations of training technique and partition choice are outside the scope of this work; our empirical evaluation is instead focused on comprehensively evaluating two types of models using a large number of random data splits across a wide range of different training set sizes.

In Fig.~\ref{fig_test_learning_curves}, the mean test accuracy is plotted versus training set size, with error bars indicating standard deviations.
The top plot shows the full model and restricted model results for the topology classification task, and the bottom plot shows the full model and restricted model results for the metallicity classification task.
Analogous plots for the train accuracy versus training set size are shown in the Supplementary Information.

For both tasks, the full model has greater mean train accuracy than the restricted model at all training set sizes, which is consistent with the fact that the full model is more expressive.
Additionally, for both types of models on both tasks, the mean training accuracy tends to increase as the training set size is decreased, whereas the mean test accuracy tends to decrease as the training set size is decreased.
This is consistent with the expectation that the models will tend to overfit more as the training set size is decreased.

\begin{figure*}[!htb]
    \centering
    \includegraphics[scale=0.6]{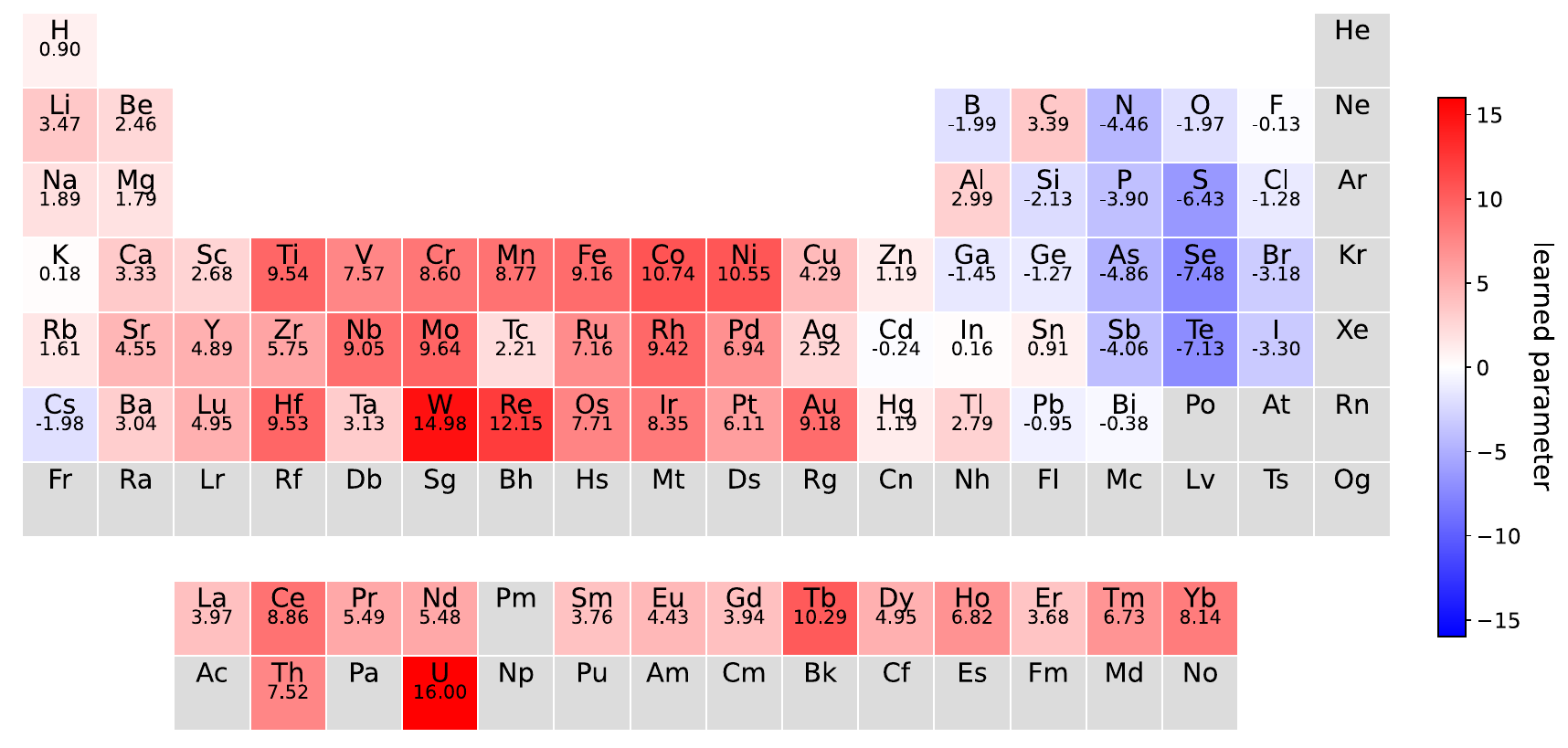}
    \caption{\textbf{Interpretable visualization of the metallicity classification model.}  The learned parameters of the full model when fit on the entire metallicity dataset are shown on the periodic table.  The values of the parameters are shown numerically and displayed visually via the color of the element.  Using this table, one can immediately perform a heuristic diagnosis of whether a material is a metal by looking at the sign of the appropriately weighted average of the material's constituent elements' parameters. \label{fig_periodic_table_visualization_of_parameters}}
\end{figure*}

For both tasks, the full model has greater mean test accuracy than the restricted model for large training set sizes.
This is also consistent with intuition -- given sufficient training data, the more expressive model should have better test performance.
Note, though, that the gap between full model test accuracy and restricted model test accuracy in the high training data regime is considerably smaller for the topology task than for the metallicity task.
The fact that restricting the hypothesis class (and reducing the number of parameters) by using the restricted model actually does not lead to much performance deterioration for topology classification in the high training data regime may shed some light on the amount of information that is actually needed for heuristic topology diagnosis.

The two tasks exhibit somewhat different behavior for smaller training set sizes.
For the topology task, the restricted model has greater mean test accuracy than the full model for all training set sizes below a certain threshold (as the training set size is shrunk, once the two mean test accuracy curves cross, they do not cross again).
For the metallicity task, the restricted model has greater mean test accuracy than the full model for some training set sizes, but has lower mean test accuracy than the full model for the smallest training set sizes (as the training set size is shrunk, the two mean test accuracy curves cross twice).
Note, however, that for both the topology task and the metallicity task, the gaps in mean test accuracy between the full model and restricted model at the smallest training set sizes are quite small in comparison to the associated standard deviations.
Collectively, these empirical results suggest that the restricted model does not require as much training data to obtain a reasonably good level of test accuracy (depending on what one considers to be reasonably good).
For example, in both of the plots in Fig.~\ref{fig_test_learning_curves}, the red curve (corresponding to the restricted model's mean test accuracy) crosses the $75 \%$ threshold at a considerably smaller training set size than the blue curve (corresponding to the full model's mean test accuracy).

\subsection{Model Visualization \label{subsection_model_visualization}}

As we see from the results in Section~\ref{subsection_performance_evaluation}, for both the topology task and the metallicity task, it is better from a predictive accuracy point of view to use the full model in the high training data regime.
We can visualize the chemical intuition learned by the full model by displaying each learned parameter $t_E$ on the periodic table.
Periodic table visualizations of full models for the topology task are shown in Fig. 2 and Fig. S4 of ref.~\cite{ma2023topogivity} (note that the method for learning parameters from data used in this paper is more similar to the method that was used to generate Fig. S4 of ref.~\cite{ma2023topogivity} than to the method that was used to generate Fig. 2 of ref.~\cite{ma2023topogivity}).

A periodic table visualization of the full model for the metallicity task is shown in Fig.~\ref{fig_periodic_table_visualization_of_parameters}.
These parameter values correspond to fitting a full model on the entire metallicity dataset using the model implementation and training approach described in Section~\ref{subsection_model_implementation_and_training}.
The collection of learned parameters captures chemical intuition in the sense that each element's parameter loosely captures the tendency of the element to form metals (a greater parameter value loosely corresponds to a greater tendency).
Note that this is about the tendency of the element to form a metallic compound when put together with other elements, as opposed to whether the element is metallic as a pure element.
For example, this table indicates that Cesium tends to form compounds that are not metallic when combined with other elements (whereas we know that pure Cesium is metallic).
See Supplementary Information for further details.

For a given material, one can look up the learned parameter values of its constituent elements in the table in Fig.~\ref{fig_periodic_table_visualization_of_parameters} and obtain the heuristic diagnosis of whether it is a metal by taking the weighted average with respect to the chemical formula subscripts.
For example, for the material $\mathrm{Li Ca_2 Mg}$ that was shown in Fig.~\ref{fig_overview_schematic_of_modeling_framework}, the weighted average is $\frac{1}{4} t_{\mathrm{Li}} + \frac{1}{2}t_\mathrm{Ca} + \frac{1}{4} t_{\mathrm{Mg}} = \frac{1}{4}(3.47) + \frac{1}{2}(3.33) + \frac{1}{4} (1.79) = 2.98$; since this number is positive, the heuristic diagnosis is that this material is a metal, which is correct.
As another example, we can consider the well-known semiconductor GaAs; the weighted average for this material is $\frac{1}{2}t_{\mathrm{Ga}} + \frac{1}{2} t_{\mathrm{As}} = \frac{1}{2}(-1.45) + \frac{1}{2}(-4.86) = -3.16$, which is negative and so the heuristic diagnosis is that this is not a metal, which is correct.

Recall that the restricted model can be viewed as a model that builds in periodic table structure by tying together weights for elements in the same set of the partition; recall also that the choice of partition for our empirical studies of the restricted model was based on a set for each column of the periodic table (that appears in the dataset) as well as one set for the f-block elements.
In contrast, the full model has a separate parameter for each element and does not explicitly build in any knowledge of the periodic table.
Nevertheless, we can see from the visualization of the full model in Fig.~\ref{fig_periodic_table_visualization_of_parameters} that some of the periodic table structure has been directly learned from the data.
For example, this can be seen by looking at the signs of the parameters.
Out of the 17 periodic table columns that do appear in this dataset (column 18 does not appear), 13 of them consist entirely of elements whose learned parameter values have the same sign.
Additionally, all of the f-block elements have learned parameter values with the same sign.

\section{Discussion \label{section_discussion}}

In this paper, we expanded on the ML approach introduced in our topogivity work~\cite{ma2023topogivity} both in terms of methodological framework and in terms of problem setting.
Methodologically, we introduced a framework for incorporating chemistry-informed inductive bias by partitioning elements so that similar elements are grouped together, and we studied it empirically for a choice of partition based on the periodic table structure.
In terms of problem setting, we considered another task in addition to topology classification -- namely metallicity classification.

Our empirical results indicate that the simple heuristic rule style of ML modeling -- which we originally introduced for topology classification~\cite{ma2023topogivity} -- also works well for metallicity classification.
In particular, when using 4,000 training samples, the full model obtained a test accuracy of $86.4 \pm 1.7 \%$ on the metallicity classification task.
While a complex black box model would still provide greater predictive accuracy, the strength of the simple heuristic rule approach lies in the fact that it is highly interpretable (as illustrated by the periodic table visualization in Fig.~\ref{fig_periodic_table_visualization_of_parameters}) while still obtaining reasonably good accuracy.

The development of ML methods for data-scarce applications is an ongoing active research topic in materials science~\cite{xu2023small,ryan2023prospective,gupta2021cross,yamada2019predicting,jha2019enhancing,ju2021exploring,zhang2018strategy,chang2022towards,ouyang2019simultaneous,feng2021general}.
For the metallicity classification and topology classification tasks, there actually is a sufficient amount of data in the real world, and so one should use the full model (which we empirically observed to have greater test accuracy in the high training data regime).
Nevertheless, our extensive characterization of performance versus training set size (by artificially restricting the training set size) for these two tasks can provide a guide for which type of model -- the full model with one parameter per element or the restricted model with chemistry-informed parameter tying -- would be more suitable if presented with a new task with a given (maybe limited) training set size.
For each task we examined here, there are some training set sizes where the full model has better test accuracy and other training set sizes where the restricted model has better test accuracy.

While the full model is already quite simple, the restricted model is even simpler.
In the opposite direction, an intriguing avenue for future work is the development of models that are somewhat more complex than the full model, yet still partially retain the thematic ideas that enable its chemical interpretability -- for example, by introducing parameters associated with the simultaneous presence of specific pairs of elements.
Positioning these more complex models alongside the full and restricted models could help illuminate the important trade-off between interpretability and accuracy in the broader landscape of ML for materials.


\FloatBarrier

\section*{Acknowledgements}

We thank Yang Zhang, Thomas Christensen, Hoi Chun Po, Li Jing, and Liang Fu for providing insight and for collaborating on the related work in which we introduced the topogivity model~\cite{ma2023topogivity}.
We also appreciate valuable comments and suggestions from Edward Zhang, Rumen Dangovski, Peter Lu, Samuel Kim, Pawan Goyal, Victoria Zhang, Ali Ghorashi, Sachin Vaidya, Lindley Winslow, and Tess Smidt.
Additionally, we thank Xiangang Wan and Feng Tang for sharing their dataset of symmetry-based topological classification of materials~\cite{tang2019comprehensive}.

A.M. acknowledges support both from the National Science Foundation Graduate Research Fellowship under Grant No. 1745302 as well as from the MIT EECS Alan L. McWhorter Fellowship.
This work is supported in part by the U. S. Army Research Office through the Institute for Soldier Nanotechnologies at MIT, under Collaborative Agreement Number W911NF-23-2-0121.
This work is also supported in part by the U.S. Office of Naval Research (ONR) Multidisciplinary University Research Initiative (MURI) Grant No. N00014-20-1-2325 on Robust Photonic Materials with High-Order Topological Protection.
This work is also supported in part by the National Science Foundation under Cooperative Agreement PHY-2019786 (The NSF AI Institute for Artificial Intelligence and Fundamental Interactions, \url{http://iaifi.org/}).
Research was sponsored in part by the Department of the Air Force Artificial Intelligence Accelerator and was accomplished under Cooperative Agreement Number FA8750-19-2-1000. The views and conclusions contained in this document are those of the authors and should not be interpreted as representing the official policies, either expressed or implied, of the Department of the Air Force or the U.S. Government. The U.S. Government is authorized to reproduce and distribute reprints for Government purposes notwithstanding any copyright notation herein.

\section*{Code availability}

The code that underlies the results in this paper will be made available in a public repository.

\def\bibsection{\section*{\refname}}
\bibliographystyle{apsrev4-2-longbib}
\bibliography{refs}

\end{document}